# Emergent risks in the Mt. Everest region in the time of anthropogenic climate change

Kimberley R. Miner (1,2)[*], Paul A. Mayewski (1), Saraju K. Baidya (11), Kenneth Broad (10,12,14), Heather Clifford (1,9), Ananta P. Gajurel (4), Bibek Giri (3), Mary Hubbard (3), Corey Jaskolski (14.10), Heather Koldewey (8), Wei Li (13), Tom Matthews (5), Imogen Napper (7), Baker Perry (6), Mariusz Potocki (1,9), John C. Priscu (13), Alex Tait (10), Richard Thompson (7), Subash Tuladhar (11)

1. Climate Change Institute, University of Maine, USA
2. Jet Propulsion Lab, California Institute of Technology, USA
3. Department of Earth Sciences, Montana State University, USA
4. Dept. of Geology, Tribhuvan University, Nepal
5. Dept. of Geography and Environment, Loughborough University, UK
6. Dept. of Geography and Planning, Appalachian State University, USA
7. International Marine Litter Research Unit, University of Plymouth, UK
8. Zoological Society of London, London, U.K.
9. School of Earth and Climate Sciences, University of Maine, USA
10. National Geographic Society, Washington DC, USA
11. Department of Hydrology and Meteorology, Kathmandu, Nepal
12. Abess Center for Ecosystem Science and Policy, University of Miami, USA
13. Dept. of Land Resources and Environmental Sciences, Montana State University, USA
14. Virtual Wonders, LLC, Wisconsin, USA

*Further information and requests for resources and reagents should be directed to and will be fulfilled by the Lead Contact and Corresponding Author: Kimberley.Miner@maine.edu

**Summary**

In April and May 2019, as part of National Geographic and Rolex's Perpetual Planet Everest Expedition (hereafter 2019 Everest Expedition), the most interdisciplinary scientific effort ever launched on the mountain was conducted. This research identified many changing dynamics, including emergent risks resulting from natural and anthropogenic changes to the natural system. We have identified compounded risks to ecosystem and human health, geologic hazards, and changing climbing conditions that can affect the local community, climbers, and trekkers in the future. This review brings together perspectives from across the biological, geological, and health sciences to better understand emergent risks on Mt. Everest and in the Khumbu region. Understanding and mitigating these risks is critical for the ~10,000 people living in the Khumbu region, as well as the thousands of visiting trekkers and the hundreds of climbers who attempt to summit each year.

**Introduction**

Mountain systems are changing rapidly throughout the world in response to climate change and human intervention. In particular, glaciers in the Hindu Kush-Himalaya are decreasing in extent and volume at dramatic rates with significant consequences for water availability, hazards (glacier outburst floods, slope failure), ecosystems, and socio-economic futures.[112] Local increases in the human footprint of the Khumbu Valley add additional socio-economic and human and ecosystem health stresses. The 250 million inhabitants of the Himalaya region and the many thousands of tourists, trekkers, and climbers who visit every year are not always aware of the rapid pace of ecological change. For all of the people and ecosystems in the Himalaya, it is critically important to clarify potential risks, mitigation, and plans for adaptation. However, research into global change dynamics is needed in mountainous regions above 5000m, and the Himalaya contains the majority of the world's mountains above 8000m.

The 2019 Everest Expedition sought to complete novel, interdisciplinary research at and above 5,000m, including biology, geology, glaciology, mapping, and meteorology (Mayewski, this issue). The majority of data presented in this study was collected during the 2019 pre-monsoon season,[30,] so it represents a unique, yet still partial, snapshot. Here, we use the new data, combined with existing research, to highlight existing and emerging risks on Mt. Everest and the watershed below.

**Physical Environment**

*Introduction to geologic risks*

The spectacular high elevation mountains and topographic relief in the Himalaya are the product of the balance between crustal deformation in a convergent tectonic setting and the impacts of erosional processes dissecting the Earth's crust in this region.[1,2] Geological and hydro-meteorological (both the tectonic and erosional) processes become hazards in the Khumbu region as earthquakes, landslides, and flooding events.[4] As the Indian continent continues its generally northward movement with respect to the Asian continent, this interplay of processes shapes the landscape of this mountain belt. As a result, hillslopes tend to be steep, and rates of erosion are high, calibrated by the strength of the material being eroded, topography, vegetation, and extreme climatic conditions.[1,2]

In the Khumbu region of eastern Nepal, the rocks predominantly consist of gneiss, schist, quartzite, marble, and leucogranitic intrusions.[5] Given the history of faulting in the area, an abundance of precipitation, particularly during the monsoon, and freeze-thaw cycles, these rocks are prone to physical weathering through fracturing (e.g., ice wedging).[6] The steep hillslopes are then susceptible to erosion via gravitationally driven events that range from slow and small-volume, to fast and high-volume movements.[2] It is these high-volume, high-velocity landslides, including rockfalls, rockslides, and debris flows, that pose a geologic hazard to people and property in the Khumbu region.[1,7-8] Indeed, it has been documented that earthquakes often trigger landslides and avalanches of all sizes.[9] Understanding the glacial and geomorphological changes is critical to protecting both the climbers and villages of the Khumbu region.

*Landslide risks*

The landscape of the Khumbu region is shaped by glacial and fluvial erosion spanning from the last ice age to the present day. There is a diversity of landslide scarps and deposits, and the village of Namche Bazar (below Mt. Everest) was built in the bowl-shaped scarp from a past landslide.[11] That scarp cuts into an earlier massive rockslide, the Khumjung slide, which is currently the sixth-largest rockslide documented in Nepal.[12] This massive rockslide is of unknown age but resulted from a $2.1 \times 10^9$ m$^3$ volume of rock material sliding southward off the Khumbila peak, where remnants make up the small E-W trending ridge that separates Khumjung and Kunde from Namche.[2] Rockslides of this magnitude are thought to account for a large percentage of the erosion across the high-relief Himalaya.[11-12] High-magnitude slides occur infrequently and may be triggered by earthquake events,[12-13] where rock strength plays a role in the frequency of occurrence.

Rivers and Glaciers are erosional transport agents for rock material, breaking down bedrock outcrops and moving eroded material down the valley.[14] The Khumbu region typically has headwalls on the order of ~55°, the angle of which promotes large-scale rockfall events (catastrophic rock failure) in the highest parts of the range. As climate changes and glaciers retreat, they melt away from the headwalls, exposing steep rock faces and increasing their vulnerability to failure. A time series of remote sensing imagery from 1962 to 2019 reveals considerable ice loss in the Khumbu region, particularly the Imja and Baruntse valley glacier systems, with newly exposed rock risking catastrophic failure (Bolch et al., this issue). All landslides, including those triggered by earthquakes and glacial recession, have the potential to introduce cascading hazards, including the damming of river valleys, rockslides, avalanches, and flood events[1-3,15].

*Earthquake risks*
Rocks of the Khumbu region have been subjected to deformational processes for the past ~55 million years, creating regular planes of failure. These planes can be due to a weaker lithology or prior faulting or shearing of the bedrock.[16] In the oldest examples of deformation in the Greater Himalayan Sequence, deep rocks were recrystallized at the mineral grain-scale during and after deformation.[17] Younger episodes of deformation have developed planes of weakness through alignment of platy minerals or the fracturing or pulverizing of rock material. Once hillslopes are steepened by the incision of streams and glaciers, these weak planes are vulnerable to failure.[2,16] The Khumbu region has several young faults that enhance the risk of a mass movement, including rock movement such as E-W striking thrust structures,[3] E-W striking extensional structures,[6,18] NE striking strike-slip structures.[19,20]

Active faults at depth exacerbate landslide risk through earthquake occurrence.[13,21] Following the 2015 ($M_w$ 7.8) Gorkha earthquake in Nepal, a debris avalanche consisting of ice, snow, and rock material buried the Langtang village, killing more than 350 people.[22,9] The earthquake caused a similar massive snow and ice avalanche at basecamp and killed more than 20 people.[23] Kargel et al. (2016) mapped more than 4300 landslides triggered by that earthquake, including throughout the Khumbu and Dudh Kosi valley.[7,9]

There are too many variables in our Earth's heterogeneous crust to be able to predict earthquakes precisely. Still, seismologists have done significant work studying past earthquakes and rates of modern tectonic plate movement. This research allows scientists to understand areas along the Himalaya that are overdue for slip events and at higher risk of earthquakes.[24,1,8] Like all other parts of the Himalaya, the Khumbu is susceptible to future high-magnitude earthquakes and the significant risks to humans and wildlife those events could cause. Seismic events may increase the probability of mass movement of both rock and ice, highlighting the interconnectivity of high mountain hazards in tectonically active mountain belts.

*Glacial lake outburst flooding risks*

Glacial lake outburst floods are another geologic hazard that relates to the evolving geomorphology under a warming climate. Outburst floods occur as glaciers retreat and leave newly formed lakes impounded by their former moraine walls. These lakes risk catastrophic breakthrough of this impoundment and subsequent flooding. Outburst floods of glacial lakes are most often triggered by mass movements into the lakes,[25] including avalanches, landslides or rockfall (Figure 1).[26,10] Evidence from sediment coring of proglacial Lake Gokyo, demonstrates that mass movements into lakes have occurred, even at high elevations (4,750m; Gajurel et al., this issue). As glacial melt accelerates, the formation of supraglacial lakes on the glacier surface also becomes a hazard (Figure 2)[27] as these lakes are isolated from meltwater release points. If supraglacial lakes suddenly break their banks, they can release significant water flow and downstream flooding[15].

To assess the risk posed by outburst floods, detailed digital elevation data are used to map glacier surfaces and increase the understanding of the terminal and supraglacial lakes.[28] New decimeter level data collected for the entire Khumbu glacier by our field team enables further evaluation of glacial lake development and its potential to threaten downstream communities (Tait et al., in review). Though there has been a history of outburst floods in the Khumbu,[15] lakes are carefully monitored, and engineering modifications, including the lowering of Imja lake, may reduce the risk.[25,29]

The Khumbu region of Nepal, including Sagarmatha National Park, is host to several geologic hazards and contains a cryptic record of past events. This mountain region contains the highest elevation in the world, and it balances tectonic uplift with erosional processes. The result is a network of steep glacial headwalls with active erosion and stream incision. As our climate changes, we see the retreat of these glaciers, growth of glacial lakes, and heightened risk from mass movements and flooding events. The seismic hazard is ever-present throughout the Himalaya and heightens the risk of mass movements and avalanches. The highest risk areas are often concentrated along the river valleys or steep valley walls, locations that also host most of the settlements in the Mt. Everest watershed and beyond, enhancing the potential for human impacts.

*Hydrological risks*
Superimposed on the geological risks of landslides, earthquakes, and glacial lake outburst floods, hydrological conditions can also become hazardous in a changing climate. However, high-elevation continuous hydrological monitoring is scarce. Our team installed a network of five automatic weather stations (AWSs) to investigate water resources and temperature in the context of climate change and to improve climber safety.[30] Observations indicate exceptionally high values of incoming solar radiation that can cause surface melting at over 8,000m even with ambient air temperatures well below freezing.[30] This widespread melt may, therefore, be mobilizing PFAS, microplastics, and other hazardous substances in the Everest ecosystem (Miner et al., this issue; Napper et al., this issue; Clifford et al., this issue).
Throughout the Khumbu, high-volume rain events have triggered dangerous flooding and secondary landslides.[32] The likelihood of these hazards may increase in the region, consistent with high-volume precipitation trends across portions of South Asia[33] and projections of future climate.[34] Growing flood risk is also suggested by regional increases in average river discharge projected for the coming decades.[28,40] These changes suggest an increase in the likelihood of heavy rain (as supposed to snow) falling in glacierized basins, which may lead to increased flood risk, especially if intense melt and rain events build on one another to increase the long-term risk to the region.

Although heavy snowstorms occur throughout the year above ~5,500 m in the Nepal Himalaya, significant snowfall associated with westerly disturbances also occurs at lower elevations (>3,000 m).[40] Major storms may deposit up to 100mm liquid-equivalent precipitation over a multi-day period at elevations of 3,000 – 4,000m in the Khumbu region (Perry et al., this issue). Snowstorms present numerous travel difficulties and substantially increase avalanche hazards across the region, even at lower elevations (Figure 1). In particular, heavy snow associated with landfalling tropical cyclones poses a significant and increasing threat to the broader Himalayan region, as evidenced by the deadly October 2014 snowstorm in the Annapurna region associated with the remnants of tropical cyclone Hudhud.[41]

However, the corollary risk is also present. Average river flows may decline in the latter half of the century if glacier retreat is sustained.[40] Subsequent water stress in the Everest region could become common as the buffering effect from glacier meltwater against seasonal meteorological drought is reduced.[41] In turn, this could challenge the food and economic security of local communities, as the regularity of rain needed to maintain crops is altered.[42]

*Atmospheric Risks*

Climbers who venture to the upper slopes of Everest are at particular risk of extreme cold. When combined with strong winds, facial frostbite can be likely within as little as one minute during winter storms, and approximately seven minutes during the popular spring climbing season.[43,44] High winds also create a mechanical risk as anecdotal accounts of mountaineers being blown from the mountain as high wind events increase danger at elevation.[31] Our analyses suggest that the freezing level during the monsoon (JJAS) has risen by 2.1 m yr$^{-1}$ (1979-2019; Perry et al., this issue). Higher freezing levels can be a secondary risk from tropical cyclones from the Bay of Bengal,[40] increasing the number of risks to climbers introduced by one storm event.

Ozone uptake by humans can cause a range of health impacts, including acute effects on otherwise healthy adults. The most severe impacts include lung inflammation and damage due to oxidative damage to cells in the airways.[45,46,47] However, almost 90% of ozone is found in the stratosphere, where it is formed by photolysis of molecular oxygen, and plays an essential role in absorbing UV-B solar radiation.[48] Ozone produced in the stratosphere can enter the troposphere via the slow overturning of the Brewer-Dobson circulation,[49] and through transient features of the atmospheric flow, primarily tropospheric folds.[50]

Research from the Khumbu Valley shows that ozone levels regularly exceed recommended safety limits.[51] This is primarily due to the long-range transport of polluted tropospheric air masses from the Indo-Gangetic Plain, and intrusions of stratospheric air.[52,53,54,55] Intermittent measurements from high altitudes on Mt. Everest suggest that levels of ozone generally increase with elevation.[44,56,57] This is consistent with the increased influence of stratospheric air and accelerated phytochemistry from enhanced solar radiation at higher altitudes.[55,58] We address the limitations of the intermittent measurement record continuity by estimating ozone concentrations on Mount Everest using the ERA5 reanalysis.[59] Our results confirm that ozone increases with elevation, and highlights that ozone concentrations peak in the spring, in agreement with previous investigations in the region.[51,60,55] This timing is consistent with the seasonality in tropospheric ozone levels, and the frequent springtime incursions of polluted air from the Indo-Gangetic Plains.[52] The spring peak may also be partly attributable to the intrusions of stratospheric air masses that are common during this period. [55,58]

The timing and location of these peak ozone concentrations may place mountaineers at significant risk. May is the most popular month for Everest summit attempts,[61,] and this is also the month when ozone safety levels identified by the World Health Organisation (WHO) are breached most frequently. Climbers attempting to summit Everest may commonly encounter ozone concentrations at levels that are associated with measurable reductions to lung function in the short term (80 ppb), or significant health impacts in the long term (120 ppb).[46] Combined with the analysis of high-frequency changes in oxygen availability by Matthews et al. (this issue), our ozone assessment highlights that variability in the atmospheric composition may substantially influence the risk for Everest climbers. Forecasts of air quality near the summit could, therefore, be developed to mediate risk, particularly for those attempting to climb without supplementary oxygen.

**Pollution and Pathogens**

*Legacy chemical pollution risks*

Throughout the Northern Hemisphere, centuries-old ice stored in glaciers is melting to reveal the chemical footprint of human chemical use.[62] Studies throughout Europe, Asia, and North America have identified chemicals as diverse as pesticides, industrial plasticizers, and lead in snow and glacier ice.[62] The melting of this ice has led to the contamination of water sources by both natural metals and anthropogenic in some of the world's most important water towers.[63] On Mt. Everest, research by Miner et al. (this issue) has uncovered the signature of generations of trekkers, in the form of directly deposited per-fluoroalkyl substances, or PFAS, compounds used to coat plastics for waterproofing. These compounds have been used widely since the 1950s and are ubiquitous in outdoor gear as varied as coats, boots, and tents (Miner et al., this issue). However, the deposition of PFAS into the Everest ecosystem has left a relatively high concentration across the mountain landscape. Indeed, at the levels detected, human consumption of the PFAS in Basecamp meltwater could be hazardous, with the levels in samples at or just below unacceptable levels of a few parts-per-trillion for states in the US (Miner et al., this issue). All of the meltwater utilized by climbers and guides at basecamp is collected from the local ecosystem and only treated for microbes, leading to direct uptake of PFAS in the watershed.

PFAS are only one compound found in quantity across the mountain. Our research has also uncovered DDT, Lindane, and lead. All of these compounds were found in amounts just below the screening levels for health impacts.[64] However, in combination, they could pose an increased risk to trekkers and the local community. While health impacts from one-year of water consumption may be minimal, the regular uptake of pollutants by Sherpas and guides could increase risk significantly through bioaccumulation.[64] In future research, sampling for toxic chemicals and metals must be prioritized, with researchers likely to find many more compounds left by humans across the Khumbu region's landscape.

*Metal pollution risks*

Asia is the most significant global contributor of anthropogenic trace metals to the atmosphere due to the increase in industrial production and the lack of emission controls in this region over the last century[65]. Some trace metals (e.g., Pb, As, Cd, Cr) are classified as systemic toxins and can cause adverse health effects, including organ damage, cardiovascular disease, developmental abnormalities, cancer, and neurological disorders[72]. Long-range and local transport of atmospheric trace metals can occur naturally[71,72], but are significantly amplified by macro-scale human activities (e.g., fossil fuel combustion, metal production, waste production) and local activities (e.g., agriculture, expansion in land use, biomass burning)[66-71]. Metal contamination in water used for drinking, irrigation, and ecological purposes, can create health problems if metals exceed safety levels[73]. In addition, trace metals and black soot, accumulating on glaciers, can result in glacier outburst floods and glacier volume loss by decreasing surface albedo, allowing for increased snow and ice melt[74,75].

Water and snow sampling during the 2019 Everest Expedition revealed recent increases in the atmospheric metal deposition in comparison to pre-modern ice core records[75] (Clifford et al. This Issue). We find elevated pollution-based metal and other aerosol concentrations in surface snow, with specific increases near to local villages compared to more remote sample sites. The correlation between location and concentration implies that the influx of metals may be from local activities.[77] Clifford et al. (This Issue) also found high concentrations of metals in the local glacial stream system, sourced from the Khumbu glacier. These metals are particularly problematic to the local population since glacier melt accounts for an average of 65 % of the domestic water resources during the dry, pre-monsoon season.[77] Among the toxic metals detected, several snow and stream samples contained concentrations of lead above WHO safety level guidelines[73].

With the expansion of tourism and infrastructure development (Miner et al., this issue), increases in local emissions, as well as long-range transport, could raise metal concentrations in surface snow and glacier melt streams in the Khumbu region. An immediate step to reduce the risk of exposure for the local population is to use a filter capable of removing heavy metals (e.g., activated carbon) on stream and snowmelt before drinking. Long-term management could include increased water testing throughout the below glacier watershed, and expanding local regulations on tourism, biomass emissions, and waste removal[71].

*Plastic pollution risks*

Nepal's tourism business and livelihood diversification have supported the local economy in the Khumbu region.[78,87] However, with this intensity of tourism, the potential for conflict between maintaining a healthy natural environment and development, also increases.[84] The first climbers to Mt. Everest began a century ago, and 10 years later, the South Col (~8,000 m) on Mt. Everest was described as 'the world's highest junkyard.'[79,80] More recently, the field team from this expedition found a plethora of items such as plastic bottles, oxygen bottles, food wrappers, food waste, and cigarette butts (Napper, This issue).

In recent decades, innovation and application of new plastics led to the introduction of lightweight, technical clothing and equipment, making the mountain more accessible.[83] Unfortunately, to date, a large proportion of the accumulating waste is plastic, due to these lightweight, strong, inexpensive, durable, and corrosion-resistant properties.

It is clear that as the number of people visiting Mt Everest has increased, this has resulted in associated increased accumulation of anthropogenic waste, including litter along trekking trails and plastic build-up in camps (Miner et al., this issue).[84,107] However, there has recently been a series of positive actions to address the risks of the waste situation. For example, in February 2019, China closed its Everest base camp on the Qinghai-Tibetan Plateau due to waste accumulation.[81,85-86] The government also banned single-use plastics starting January 2020, in a bid to cut down on waste left by climbers.[82] For example, in February 2019, China closed its Everest base camp on the Qinghai-Tibetan Plateau due to waste accumulation.[81,85-86] The government also banned single-use plastics starting January 2020, in a bid to cut down on waste left by climbers.[82]   According to recent legislation, the Government of Nepal have also introduced a provision for garbage management in its Mountaineering Expedition Rules, 2002 that the deposit of $4000 will be returned only after the submission of evidence of garbage management when the mountaineers descended from Mount Everest.

Given the fragile environment, extreme weather conditions, and limited infrastructure, the waste produced is often beyond local capacity to handle, increasing the stress to local towns and economies.[83] New waste management systems are required, as the complete elimination of waste locally would be extremely challenging. Plastic pollution joins the growing chemical load on the mountain as an additional challenge to local populations unable to manage the challenges associated with generations' influx of trekkers.

*Pathogens pollution risks*

In addition to anthropogenic components, atmospheric aerosols are comprised of a wide range of biological materials, including bacteria, Archaea, pollen, fungi, protists, and viruses.[88-90] Some of these aerosolized species can act as plant or human pathogens.[91-93] The aerosols and associated pathogens can be transported and deposited via precipitation onto distant glacier surfaces.[94-96] As glaciers and ice sheets recede at an accelerated pace in response to climate change,[28] the potential for impact of the enhanced release of ice-entombed pathogenic organisms increases for the downstream human populations.[97] The release of pathogenic bacteria has been noted by reports of glacial transport of fecal bacteria from climbing activities on Mt. McKinley's Kahiltna Glacier, Alaska.[98,99] In this case, buried human waste from climbers emerged at the glacier surface within decades of deposition.[98,99]

We analyzed environmental DNA[100] in ice cores from Mount Everest South Col (ESC, at 8030 m asl) and Mount Everest Base Camp (EBC, at 5400 m asl) to determine the presence of potential pathogens along this popular climbing route. According to the methods outlined in Priscu et al. (this issue), bacterial operational taxonomic units (OTUs) closely related to *Clostridioides difficile* (98.5% similarity, NCBI accession NR_113132) and *Pseudomonas aeruginosa* (99.2% similarity, NCBI accession NR_117678) were identified in sequence libraries derived from ice core samples collected during the expedition. Both species have the potential to impact human health. *Clostridioides difficile* can cause toxin-mediated infections ranging from mild diarrhea to death, with the dominant transmission through the ingestion of spores.[101] *Pseudomonas aeruginosa* is found in various habitats, including soil, marshes, marines, plant, and animal tissues, and is noted for its resistance to antibiotics and disinfectants. Owing to this resistance that allows for decreased competition, *P. aeruginosa* is a primary opportunistic human pathogen known to threaten burn victims, urinary-tract infections patients, and pneumonia patients with low immune system response.[102]

OTUs closely related to *Streptococcus vestibularis* (99.2% similarity, NCBI accession NR_042777) were identified only in the highest ice core (8030m). *S. vestibularis* forms a significant part of the healthy human oral cavity but is known in association with infective endocarditis.[103] Sequences closely related to *Staphylococcus epidermidis* (99.2% similarity, NCBI accession NR_113957) were also found in the Base Camp ice core sample only. *S. epidermidis* is an opportunistic human pathogen that often infects patients with compromised immune systems.[104]

Mitochondrial 16S ribosomal-RNA genes related to two pathogenic fungi were also identified, including *Fusarium oxysporum* (NCBI accession KU158767), a fungal pathogen to numerous agricultural plants[105] and *Paecilomyces penicillatus* (MK069583), which infects other fungal species.[106] Although the source of these pathogenic strains within the ice core samples analyzed remains unclear, and results indicate that consumption of untreated glacial meltwater may lead to illness in local and downstream human, animal wildlife, and plant populations. Further research and indexing of pathogenic species are necessary, as the biological waste left by generations of climbers on the glacier continues to move downstream.

### **Conclusion**

As the highest mountain in the world, Mt. Everest is considered one of the most coveted climbs for mountaineers across the globe. However, the risks to the health and safety of the trekkers and residents alike are increasing rapidly as climate change and increased traffic change the dynamics of the mountain. The growing ecosystem changes are exposing the hidden impacts left by humans across landscapes, where our influences are seen even on the highest mountain on Earth. This frozen chemical signature of human interaction is an unfortunate side effect of innovation that becomes increasingly harmful as glaciers melt at an unprecedented rate. In the future, it will be necessary to continue to expand our understanding of the diversity of risks that the mountain poses, and to monitor the status of these changing conditions.

Assembling teams of interdisciplinary researchers to document and understand the dynamic environment of Mt. Everest is a significant step towards protecting this resource for generations. Understanding ice dynamics, geology, and precipitation dynamics are critical to begin to adapt to the changes for both the local community and the mountaineering teams. While not all of the mountaineers from Mt. Everest are from the Khumbu region, their actions while on the mountain leave a legacy that will impact the downstream residents for generations. Mitigating the impacts of mountaineering, and understanding the risks that tourism and climate change bring to the local community will provide a clear picture of a sustainable way ahead.

**Acknowledgments**
This research was conducted in partnership with the National Geographic Society, Rolex, and Tribhuvan University, with approval from all relevant agencies of the Government of Nepal. We thank the communities of the Khumbu Region, our climbing support team, Shangri-La Nepal Trek, and all support staff. Thanks to The National Geographic Perpetual Planet Everest team, including Sandra Elvin, Fae Jenks, Alexis Bahl, and Aurora Elmore, and Aerial Filmworks for the specialized technical support.
Funding provided by the National Geographic Society and Rolex.

**Author Contributions**
All authors contributed to developing and producing the content of this manuscript

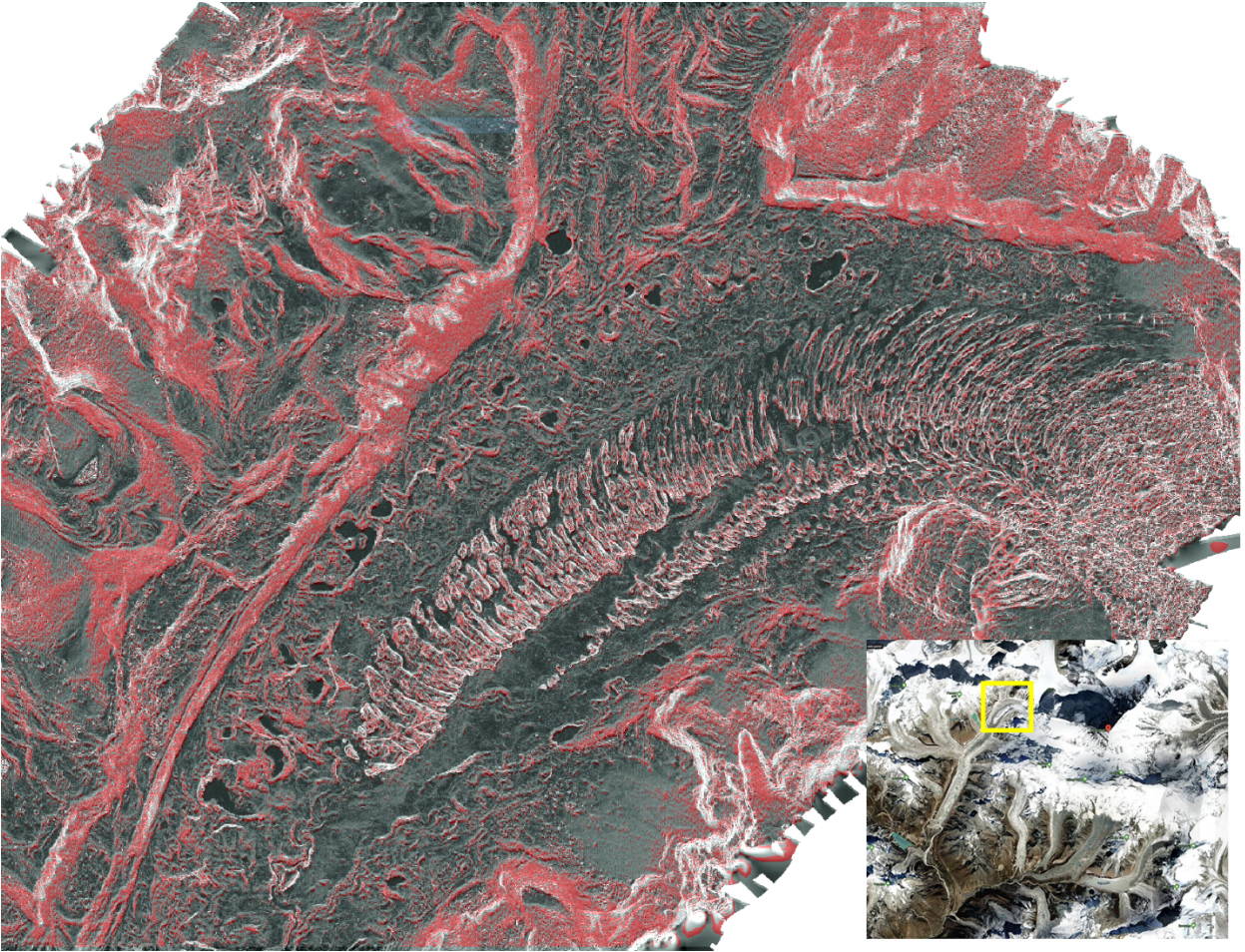

Figure 1. This slope map composite identifies areas with slopes between 30 and 45 degrees (red shading), the angle at which avalanches are most likely. Many of these zones and their potential runout regions are very close to basecamp and the climbing route, reiterating the risk of avalanche is always present on Everest.

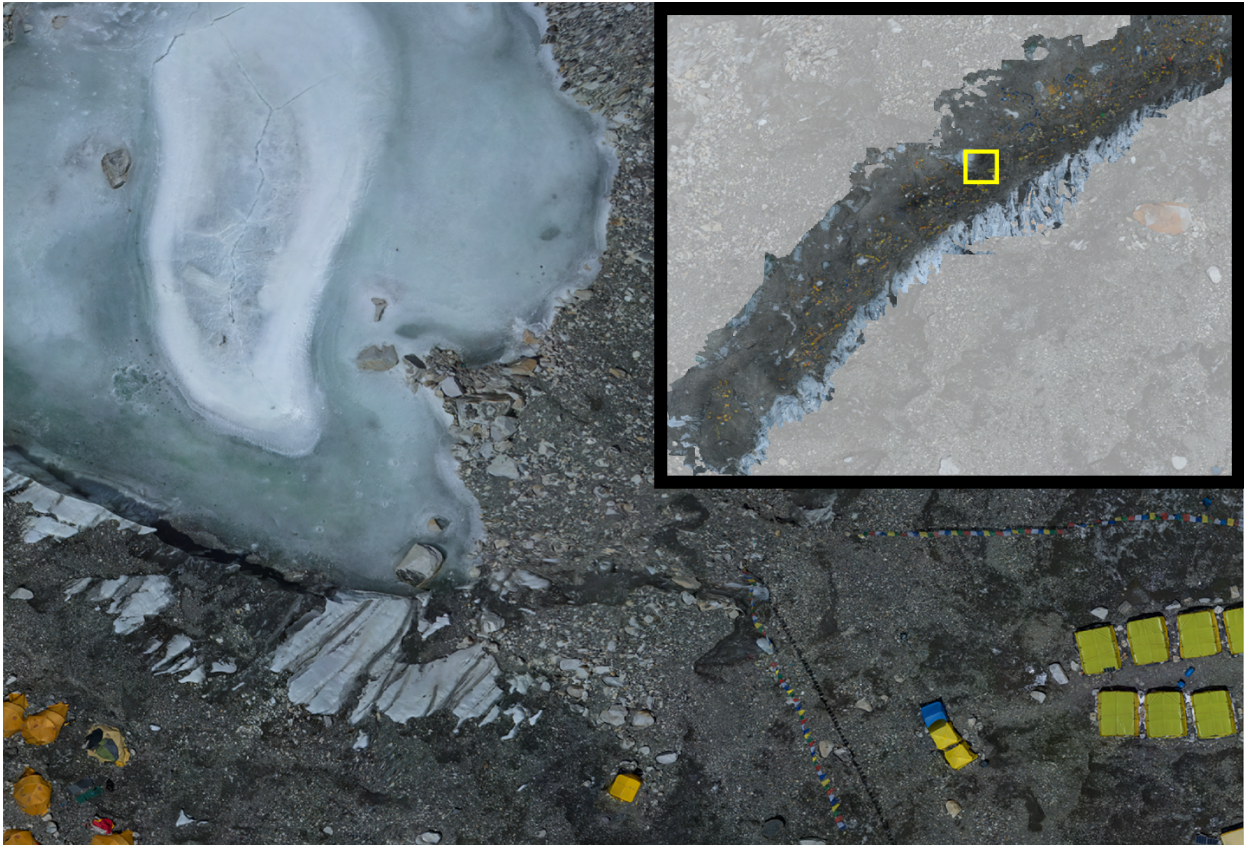

Figure 2. High-resolution 3D scans of basecamp show many instances of tents erected near melt pools as seen in this image. These melt pools hint at the amount of liquid water that may be on top of and beneath the glacier surface, creating significant risks to basecamp through possible surface collapse, supraglacial flooding and glacial outburst flooding.